\documentclass[10pt,aps,preprint]{revtex4}
\usepackage[T1]{fontenc}
\usepackage[utf8]{inputenc}
\usepackage{textcomp}
\usepackage{amsmath}
\usepackage{amssymb}

\makeatletter

\newcommand{\lyxmathsym}[1]{\ifmmode\begingroup\def\b@ld{bold}
  \text{\ifx\math@version\b@ld\bfseries\fi#1}\endgroup\else#1\fi}

\providecommand{\tabularnewline}{\\}

\@ifundefined{textcolor}{}
{%
 \definecolor{BLACK}{gray}{0}
 \definecolor{WHITE}{gray}{1}
 \definecolor{RED}{rgb}{1,0,0}
 \definecolor{GREEN}{rgb}{0,1,0}
 \definecolor{BLUE}{rgb}{0,0,1}
 \definecolor{CYAN}{cmyk}{1,0,0,0}
 \definecolor{MAGENTA}{cmyk}{0,1,0,0}
 \definecolor{YELLOW}{cmyk}{0,0,1,0}
}

\makeatother

\begin{document}

\title{Hybrid meson interpretation of the exotic resonance $\pi_{1}(1600)$ }

\author{Azzeddine Benhamida }

\affiliation{LPTO, University of Oran 1-Ahmed Ben Bella}

\email{benhamida.azzeddine@edu.univ-oran1.dz, benhmidaazou@gmail.com}

\author{Lahouari Semlala}

\affiliation{Ecole Supérieure en Génie Electrique et Energétique d'Oran}

\email{semlala.lahouari@univ-oran.dz, lsemlala@yahoo.fr}

\begin{abstract}
The exotic J$^{PC}=$1$^{-+}$resonance $\pi_{1}(1600)$ is examined
in the framework of the Quark Model with Constituent Gluon (QMCG).
We report the possibility of interpreting that resonance as $q\bar{q}g$
meson, with a masse $\simeq1.65_{-0.04}^{+0.05}$ GeV and a decay
width to $\rho\pi$ $\simeq0.28_{-0.09}^{+0.14}$ GeV. 
\end{abstract}
\maketitle

\section{Introduction}

Over the last four decades, intensive experimental activity has been
carried out seeking to detect new hadrons beyond the quark model:
glueballs or gluonium, hybrids, diquonia, tetraquarks,... These \textquotedbl{}exotic\textquotedbl{}
species are the most likely new hadrons allowed by the QCD, and are
the subject of numerous researches, both theoretical and experimental.

Hybrid mesons (quark-antiquark-gluon) can have $J^{PC}$ quantum numbers
which are not allowed by the naive quark model, like $0^{--}$, $0^{+-}$,
$1^{-+}$, $2^{+-}$, then they can not mix with the standard mesons
and hence can facilitate their observation. These ”exotic”\ objects
are the most promising new species of hadrons allowed by QCD and subject
of lot of works both in the theoretical and experimental levels. In
fact, several $J^{PC}$=$1^{-+}$ exotic resonances have been claimed
to be identified, especially $\pi_{1}(1600)$ and $\pi_{1}(1400)$
have received great interest, but some doubts are raised about the
last one, for a review, see Ref\cite{Meyer-2015}.

In the theoretical framework, these hybrid mesons were studied from
different models: lattice QCD \cite{LQCD-0,LQCD-1,LQCD-2,LQCD-3,LQCD-mass,LQCD-decay},
flux tube model \cite{Flux-0,Flux-1,Flux-1-SS-2,Flux-2}, bag model
\cite{bag-1,bag-2}, QCD sum-rules \cite{sum-1,sum-decay,sum-2,sum-3,sum-4},
constituent gluon models \cite{mandula,orsay,iddir-88,Ishida-92,Ishida-89,doctorat,habilitation,Kalashnikova}
and effective Hamiltonian model \cite{CoulmbGauge-1,CoulmbGauge-2,CoulmbGauge-3}.
Some of these models can perform both estimations of masses and decay
widths, they predicted that the lightest hybrid mesons will be in
$1.4-2.1$ GeV mass range which is consistent with the confirmed $1^{\text{-+}}$candidates.

The $\pi_{1}(1600)$ was observed decaying into $b_{1}\pi$, $f_{1}\pi$,
$\eta'\pi$ and $\rho\pi$. But although the first three modes have
been confirmed for a long time, the $\rho\pi$ mode has been incorporated
only recently in PDG since 2018 \cite{PDG2018}. Indeed, This mode
is forbidden due to the ``standard'' flux-tube predictions in a
symmetry limit where the $\rho$ and $\pi$ have the same size and
in the case where the decay is triggered by breaking the flux-tube
\cite{Flux-0,Flux-1,Flux-1-SS-2,Flux-2}, although a value of $57$
MeV was calculated beyond this limit \cite{Flux-3}. This remains
quite far from the very recent measurements made by COMPASS experiment
\cite{COMPASS-2018}, see also PDG-2018 \cite{PDG2018}.

In this work, we focus our attention on the $1^{-+}$ hybrid meson
in the context of the \textit{Quark Model }with\textit{ Constituent
Gluon} (QMCG), and we shall see that this constituent glue model gives
values of the mass and the $\rho\pi$ decay width of the lightest
$1^{-+}$ $q\bar{q}g$ quite compatible with the observed exotic candidate
$\pi_{1}(1600)$.

This paper is organized as follows. In Sec. II we briefly present
the experimental situation of the exotic $J^{PC}=1^{-+}$ $\pi_{1}(1600)$.
We give predictions of the model MQGC in Sec. III and we conclude
in Sec. IV.

\section{the experiemental status of $\pi_{1}(1600)$}

We consider here only the status of the resonance $\pi_{1}(1600)$,
for a review of the experimental situation on exotic hybrid mesons
see the Ref \cite{Meyer-2015} that we have mainly used when preparing
this section .

VES Collaboration \cite{VES-conclusion} observed a broad peak at
a mass value of $\sim1.6\mathrm{\:Ge\mathrm{V}}$ in the $\eta'\pi$,
$f_{1}(1235)\pi$, and $b_{1}(1235)\pi$ systems, interpreted as an
exotic resonance of width about $300$ MeV. Actually they are unable
to make a definitive conclusion on the resonance nature of it. For
the $\rho\pi$ final state, they unable to conclude that the $\pi_{1}(1600)$
is present, while the following experimental relationship between
the branching fractions of the $\pi_{1}(1600)$ decays is obtained
(and therefore a limit on the branching fraction of $\rho\pi$ ):
\begin{equation}
b_{1}\pi:f_{1}\pi:\rho\pi:\eta'\pi=(1.0\pm0.3):(1.1\pm0.3):<0.3:1.
\end{equation}

E852 Collaboration at BNL reported evidence for the $1^{-+}$ $\pi_{1}(1600)$
resonance decaying into $\rho\pi$ \cite{E852-adams,E852-chung},
$\eta'\pi$ \cite{E852-Ivanov}, $f_{1}\pi$ \cite{E852-Kuhn} and
$b_{1}\pi$ \cite{E852-Lu}, regarding the $\rho\pi$ channel, in
the earlier E852 analysis \cite{E852-adams,E852-chung} of $250$
K $(\pi^{-}\pi^{-}\pi^{+})$ events showed a possible evidence for
a $1^{-+}$ exotic meson with a mass of $\sim1.6$ GeV and width $\sim168$
MeV, this state have been excluded by a more refined analysis \cite{E852-Dzierba},
with $2.6$ M $(\pi^{-}\pi^{-}\pi^{+})$ and $3$ M $(\pi^{-}\pi^{0}\pi^{0})$
events.

The exotic $\pi_{1}(1600)$ is observed decaying to $b_{1}\pi$ from
the Crystal Barrel data and only results with the mass and width fixed
to the PDG values were reported \cite{CrystalBarrel}.

CLEO Collaboration found evidence for an exotic P-wave $\eta'\pi$
amplitude, which, if interpreted as a resonance, would have parameters
consistent with the $\pi_{1}(1600)$ state: with a mass of $1670\pm30\pm20$
MeV and a width of $240\pm50\pm60$ MeV \cite{CLEO-c}.

A search for exotic mesons in the $(\pi^{+}\pi^{+}\pi^{-})$ system
photoproduced by the charge exchange reaction $\gamma p\rightarrow\pi^{+}\pi^{+}\pi^{-}(n)$
was carried out by the CLAS Collaboration at Jefferson Lab. and no
evidence is shown of the exotic $\pi_{1}(1600)$ decaying to three
charged pions \cite{CLAS-c1,CLAS-c2}.

COMPAS collaboration observed the spin-exotic $\pi_{1}(1600)$ in
their partial-wave analysis of the $3\pi$ final state. They reported
the observation of the $\pi_{1}(1600)$ in the $\rho\pi$ decay mode
initially with a mass $1660\pm10_{-64}^{+0}$ MeV and width $269\pm21_{-64}^{+42}$
MeV \cite{COMPASS-1}, superseded by a mass $1.600_{-0.060}^{+0.100}$
GeV and width $0.580_{-0.230}^{+0.100}$ GeV in their recent analysis
\cite{COMPASS-2018}. COMPASS collaboration has also examined the
exclusive production of $\eta\pi$ and $\eta'\pi$ and reported that
odd partial waves, which carry non-$q\bar{q}$ quantum numbers, are
suppressed in the $\eta\pi$ system relative to the $\eta'\pi$ system.
Even though they saw the exotic $1^{-+}$ wave in $\eta'\pi$ as the
dominant wave, they were unable to confirm the resonant nature of
the signal \cite{COMPASS-3}. This has recently been improved by the
JPAC collaboration \cite{JPAC-c} which performed the first coupled-channel
analysis of the $P$ wave in the $\eta^{(')}\pi$ system measured
at COMPASS \cite{COMPASS-3} and reported a single exotic $\pi_{1}$
with mass and width are determined to be $1564\pm24\pm86$ and $492\pm54\pm102$
MeV, respectively .

In conclusion, the $\pi_{1}(1600)$ was observed decaying into $b_{1}\pi$,
$f_{1}\pi$, $\eta'\pi$ and recently confirmed for $\rho\pi$ mode
by COMPAS collaboration \cite{COMPASS-2018}, it is considered by
the PDG to be an established state \cite{PDG2018}. Table \ref{tab:exp}
shows masses and the corresponding decay widths of the $\pi_{1}(1600)$
reported by different experiments.

\section{The QMCG predictions}

The nature of the gluonic field inside the hybrid meson is not yet
be clear because the gluon plays a double role: it propagates the
interaction between color sources and, being itself colored, it undergoes
the interaction. In an attempt to achieve a clearer understanding
about the hybrid nature, two important hypothesis can be retained
from the literature. The first one consider gluonic $\deg$rees of
freedom as ``excitations”\ of the ``flux tube”\ between quark
and antiquark, which leads to the linear potential, that is familiar
from quark model (flux-tube model).

In the second one, the framework of the so called \textit{Quark Model
}with\textit{ Constituent Gluon} (QMCG) supported by this work, the
hybrid meson is considered as a QCD bound state composed of a quark-antiquark
pair and (a massive) constituent gluon which interact through a phenomenological
potential. We can adapt this scheme with the idea of confined and
confining gluons (in the Landau and Coulomb gauges and in interpolating
gauges between them) \cite{confined-confining}. Confining gluons
establish an area law behavior of the Wilson loop and the linearly
rising interquark confinement, while confined gluons do not propagate
over long distances, we can accommodate the confined (massive, constituent)
gluon in coexistence with an effective quark interaction which is
confining: more details can be found in Ref \cite{doctorat,habilitation}.

``\textit{It is important to realise that the more complicated picture
emerging for QCD in the covariant gauge can certainly accommodate
confined (but not confining) gluons in coexistence with an effective
quark interaction which is confining, however.} ''\cite{confined-confining2}.

\subsection{Ingredients of the QMGC}

The QMCG is a natural expansion of the naïve quark model where the
confined gluon within the hadron matter acquires a (constituent) mass
$m_{g}$. As for quarks, this important parameter represents a dynamical
mass which is responsible for the infrared finiteness of the gluon
propagator and the ghost dressing function observed using continuum
methods (the Shwinger-Dyson Equations) and large-volume lattice simulations
or combining continuum methods with lattice data; a more complete
presentation of the subject is given for example in Refs. \cite{gluon-ghost-review1,gluon-ghost-review2,gluon-mass-SDE,gluon-mass-SDE2}.

From the phenomenological point of view, a non-vanishing gluon mass
is welcome by diffractive phenomena \cite{diffractive} and inclusive
radiative decays of $J/\psi$ and $\Upsilon$ \cite{radiative decays}.
For the glueball states, color singlet bound states of gluons are
considered to be fairly massive, e.g. about 1.5 GeV for the lowest
$0^{++}$ and about 2 GeV for the lowest $2^{++}$ , as indicated
in lattice QCD calculations \cite{glueball-LQCD1,glueball-LQCD2,glueball-LQCD3},
a simple constituent gluon picture may be approximately obtained as
$M_{GB}\simeq2m_{g}$ for the glueball mass $M_{GB}$ .

Using continuum strong QCD one infers $m_{g}\simeq0.4-0.6$ GeV \cite{gluon-mass-SDE2}
which is consistent with the lattice results: $m_{g}\sim0.5$ GeV
\cite{gluon-mass-LQCD,gluon-mass-LQCD2}.

In the present work we fix this parameter as: 
\begin{equation}
m_{g}=0.5\pm0.1\:\textrm{GeV.}
\end{equation}

The decay parameter $\alpha_{s}$ (the effective quark–gluon vertex
coupling) is the second ingredient of the model. There are many theoretical
evidences that the QCD effective charge $\alpha_{s}$ freezes at small
momenta. Therefore the infrared finiteness of the effective charge
can be considered as one of the manifestations of the phenomenon of
dynamical gluon mass generation. Phenomenology sensitive to infrared
properties of QCD gives: $\alpha_{s}(0)\simeq0.7\pm0.3$ \cite{as1,as2,as3},
while the phenomenological evidences for the strong coupling constant
freezing in the infrared are much more numerous, as with the models
where a static potential is used to compute the hadronic spectra make
use of a frozen coupling constant at long distances, for more details
see for example the Ref. \cite{as-constant}.

The effective charge obtained within the pinch technique (PT) framework
\cite{PT1,PT2}, to be denoted by $\alpha_{PT}$ , constitutes the
most direct non-Abelian generalization of the familiar concept of
the QED effective charge. Since our decay model is obtained in the
Feynman gauge \cite{orsay}, it is natural to choose $\alpha_{s}\simeq\alpha_{PT}(0)$
corresponding to the pinch technique gluon propagator, i.e. the background
field propagator calculated in the Feynman gauge. $\alpha_{PT}(0)$
is correlated to the gluon mass $m_{g}$ \cite{alphas0-1,alphas0-2}:
\begin{equation}
\alpha_{PT}(0)\sim0.6\quad\textrm{for}\quad m_{g}\sim0.5\:\textrm{GeV}.
\end{equation}

\subsection{The hybrid bound state}

We assume that the hybrid meson is a bound state of quark-antiquark
and a constituent gluon which interact through a phenomenological
potential, precisely Coulomb plus linear potential supplemented by
spin-spin, spin-orbit and tensor correction terms. The use of relativistic
kinetics is appropriate for the study of the light flavor systems
\cite{doctorat}.

For the representation of the hybrid states the following notations
are used:

\emph{l}$_{\text{g }}$\emph{\ \ }: the relative orbital momentum
of the gluon in the \emph{q$\bar{q}$} center of mass;

$l_{q\bar{q}}$ \emph{\ }: the relative orbital momentum between
\emph{q} and \emph{$\bar{q}$};

$S{}_{q\bar{q}}$ \emph{\ }: the total quarks spin.

Considering the gluon moving in the framework of the $q\overline{q}$
pair, the Parity of the hybrid will be: 
\begin{equation}
P=\left(-\right)^{l_{q\overline{q}}+1}\cdot\left(-1\right)\cdot\left(-\right)^{l_{g}}=\left(-\right)^{l_{q\bar{q}}+l_{g}},
\end{equation}

$\left(-1\right)$ being the intrinsic parity of the gluon.

The Charge Conjugation is given by: 
\begin{equation}
C=\left(-\right)^{l_{q\bar{q}}+S_{q\bar{q}}+1}.
\end{equation}

$S_{q\bar{q}}$ can takes values 0 or 1; \textit{P} and \textit{C}
impose parity restrictions on $l_{q\bar{q}},\ $and $l_{g}$ .

For lower values of orbital excitations ( $l_{q\bar{q}}$ and $l_{g}\leqslant1$)
and parity $P=-1$, hybrid states can be built by two modes: $l_{q\bar{q}}=0$
and $l_{g}=1$ which we shall refer as \emph{the gluon-excited hybrid
}(GE hybrid), and $l_{q\bar{q}}=1$ and $l_{g}=0$ which we shall
refer as \emph{the quark-excited hybrid }(QE hybrid), see Table \ref{tab:PC}
for the case $J^{\textrm{PC}}=1^{-+}$.

In the potential model the simplest approximation is to factorise
the $q\bar{q}$-wave function with the wave function of the gluon
respective to the $q\bar{q}$ center of mass (The cluster approximation).
We shall use the following lowest-lying state $q\bar{q}$-cluster
spin-space wave function: 
\begin{eqnarray}
\Psi_{JM}^{\textrm{PC}} & (\overrightarrow{\rho},\overrightarrow{\lambda})= & \left(\left(\left(\mathbf{e_{\mu_{g}}}\otimes\psi_{l_{g}}^{m_{g}}\right)_{j_{g}M_{g}}\otimes\psi_{l_{q\bar{q}}}^{m_{q\bar{q}}}\right)_{Lm}\otimes\chi_{\mu_{q\bar{q}}}^{S_{q\bar{q}}}\right)_{JM}^{\textrm{PC}}\\
 & = & \sum_{\left(L\,l_{g}j_{g}l_{q\bar{q}}S_{q\bar{q}};\textrm{PC}\right)}\Psi_{JM;Ll_{g}j_{g}l_{q\bar{q}}S_{q\bar{q}}}(\overrightarrow{\rho},\overrightarrow{\lambda}),
\end{eqnarray}
$\mathbf{e_{\mu_{g}}}$ is the gluon polarisation, $\chi_{\mu_{q\bar{q}}}^{S_{q\bar{q}}}$
is the diquark spin representation, and the sum runs over the values
of $L,\,l_{g},\,j_{g},\,l_{q\bar{q}}\,\textrm{and}\,S_{q\bar{q}}$
excluding those not consistent with P and C and: 
\[
\Psi_{JM;Ll_{g}j_{g}l_{q\bar{q}}S_{q\bar{q}}}(\overrightarrow{\rho},\overrightarrow{\lambda})=\sum_{\left(m\,m_{g}\mu_{g}M_{g}m_{q\bar{q}}\mu_{q\bar{q}}\right)}\left\langle l_{g}m_{g}\mathbf{1\textrm{\ensuremath{\mu_{g}}}}|j_{g}M_{g}\right\rangle \left\langle l_{q\bar{q}}m_{q\bar{q}}j_{g}M_{g}|Lm\right\rangle \left\langle LmS_{q\bar{q}}\mu_{q\bar{q}}|JM\right\rangle 
\]

\begin{equation}
\times\psi_{l_{g}l_{q\bar{q}}}^{m_{g}m_{q\bar{q}}}(\overrightarrow{\rho},\overrightarrow{\lambda})\mathbf{e}_{\mu_{g}}\chi_{\mu_{q\bar{q}}}^{S_{q\bar{q}}};\label{eq:Ljg-wavefunction}
\end{equation}
here the Jacobi coordinates are introduced: 
\begin{eqnarray}
\overrightarrow{\rho} & = & \overrightarrow{r}_{\bar{q}}-\overrightarrow{r}_{q},\nonumber \\
\overrightarrow{\lambda} & = & \overrightarrow{r}_{g}-\frac{M_{q}\overrightarrow{r}_{q}+M_{\bar{q}}\overrightarrow{r}_{\bar{q}}}{M_{q}+M_{\bar{q}}}.
\end{eqnarray}
The Hamiltonian is constructed, containing a phenomenological potential
which reproduces the QCD characteristics; its expression has the mathematical
“Coulomb + Linear” form, we take into account also somme relativistic
effects i.e. spin dependent interaction terms and relativistic kinetics;
a more detailled description can be found in our previous work \cite{doctorat}.

In order to make a comparison with lattice results we note that our
$1^{-+}$ wave function (Equ.\ref{eq:Ljg-wavefunction}) is related
to the so-called TE, TM and longitudinal gluon states as follows (see
Table \ref{tab:PC}): 
\begin{eqnarray}
\Psi_{1M}^{\textrm{-+}} & = & \Psi_{j_{g}=0}^{\textrm{long}}+\Psi_{j_{g}=1}^{\textrm{TE}}+\textrm{mix}\left(\Psi_{j_{g}=2}^{\textrm{TM}},\,\Psi_{j_{g}=2}^{\textrm{long}}\right)+\textrm{mix}\left(\Psi_{j_{g}=1}^{\textrm{TM}},\,\Psi_{j_{g}=1}^{\textrm{long}}\right),\label{eq:wavefunction-long-TE-TM}
\end{eqnarray}
$\textrm{mix}\left(\varphi,\psi\right)$ means a mixture of the states
$\varphi$ and $\psi.$ The ``magnetic'' TE, ``electric'' TM and
longitudinal gluons correspond to the following hybrid states \cite{orsay}:
\begin{equation}
\Psi_{j_{g}}^{\textrm{TE}}\equiv\Psi_{JM;Ll_{g}j_{g}l_{q\bar{q}}S_{q\bar{q}}}^{\textrm{TE}}=\Psi_{JM;Ll_{g}j_{g}l_{q\bar{q}}S_{q\bar{q}}}\left|_{l_{g}=j_{g}}\right.,
\end{equation}
\begin{equation}
\Psi_{j_{g}}^{\textrm{TM}}\equiv\Psi_{JM;Ll_{g}j_{g}l_{q\bar{q}}S_{q\bar{q}}}^{\textrm{TM}}=\sqrt{\frac{j_{g}+1}{2j_{g}+1}}\Psi_{JM;Ll_{g}j_{g}l_{q\bar{q}}S_{q\bar{q}}}\left|_{l_{g}=j_{g}-1}\right.+\sqrt{\frac{j_{g}}{2j_{g}+1}}\Psi_{JM;Ll_{g}j_{g}l_{q\bar{q}}S_{q\bar{q}}}\left|_{l_{g}=j_{g}+1}\right.,
\end{equation}
\begin{equation}
\Psi_{j_{g}}^{\textrm{long}}\equiv\Psi_{JM;Ll_{g}j_{g}l_{q\bar{q}}S_{q\bar{q}}}^{\textrm{long}}=-\sqrt{\frac{j_{g}}{2j_{g}+1}}\Psi_{JM;Ll_{g}j_{g}l_{q\bar{q}}S_{q\bar{q}}}\left|_{l_{g}=j_{g}-1}\right.+\sqrt{\frac{j_{g}+1}{2j_{g}+1}}\Psi_{JM;Ll_{g}j_{g}l_{q\bar{q}}S_{q\bar{q}}}\left|_{l_{g}=j_{g}+1}\right..
\end{equation}

We can rewrite the Eq. \ref{eq:wavefunction-long-TE-TM} according
to the GE and QE hybrid modes as: 
\begin{equation}
\Psi_{1M}^{\textrm{-+}}=\psi^{\textrm{GE}}+\psi^{\textrm{QE}},
\end{equation}
where: 
\begin{eqnarray}
\psi^{\textrm{GE}} & = & \Psi_{j_{g}=0}^{\textrm{long}}+\Psi_{j_{g}=1}^{\textrm{TE}}+\textrm{mix}\left(\Psi_{j_{g}=2}^{\textrm{TM}},\,\Psi_{j_{g}=2}^{\textrm{long}}\right),\label{eq:GE}\\
\psi^{\textrm{QE}} & = & \textrm{mix}\left(\Psi_{j_{g}=1}^{\textrm{TM}},\,\Psi_{j_{g}=1}^{\textrm{long}}\right).\label{eq:QE}
\end{eqnarray}

Since our gluon is assumed to be massive the longitudinal component
must be present in Eq. \ref{eq:wavefunction-long-TE-TM}, mixed with
the TM and TE gluon modes $\left(j_{g}\neq0\right)$. This is not
true in the lattice hybrid calculations where the low-laying $1^{-+}$
states are made with the particular $j_{g}^{\textrm{P}_{g}\textrm{C}_{g}}=1^{+-}$
TE-gluon mode. Indeed, although in principal lattice constuction of
the hybrid $1^{-+}$ states involves the TM and the TE modes, in the
light sector only the last mode results are widely reported since
it gives the best and the clearest signal \cite{LQCD-0,LQCD-1,LQCD-2,LQCD-3}.
From the Table \ref{tab:PC}, we notice that the TE-gluon appears
only in the GE-hybrid and is totally absent in the QE-hybrid $1^{-+}$
state. We will come back to this issue latter.

\subsection{The hybrid decay model }

To lowest order the decay of an hybrid state A into two ordinary mesons
B and C is described by the matrix element of the Hamiltonian annihilating
a gluon and creating a quark pair (QPC model): 
\begin{equation}
\left\langle BC|H|A\right\rangle =g\,f(A,B,C)\,\left(2\pi\right)^{3}\delta_{3}\left(\vec{p_{A}}-\vec{p_{B}}-\vec{p}_{C}\right).
\end{equation}

The partial width is given by: 
\begin{equation}
\Gamma_{A\rightarrow BC}=4\alpha_{s}\left|f(A,B,C)\right|^{2}\frac{P_{B}E_{B}E_{C}}{M_{A}},
\end{equation}
where $\alpha_{s}$ represents the infrared quark–gluon vertex coupling.
For more details on the decay model see Refs. \cite{orsay,doctorat},
here we focus on the main (non relativistic) results: 
\begin{enumerate}
\item the $1^{-+}$ QE-hybrid is allowed to decay into two S-wave mesons
only (the so-called ``$S+S$ '' selection rule), 
\item the $1^{-+}$ GE-hybrid is allowed to decay into a channel with one
S-wave meson and one P-wave meson only (the so-called ``$L+S$ ''
selection rule). 
\end{enumerate}
The last selection rule is also reported in the gluonic excitation
models of hybrid where the decay to two S-wave mesons is strongly
supressed, see \cite{FT-relativistic-decay} and references therein.

In the decay model that we use, the $\eta^{(')}\pi$ decay modes are
suppressed by the non-relativistic spin conservation law although
the spatial overlap is not vanishing for the QE-hybrid mode, a full
relativistic studies shall give non vanishing answer for both QE and
GE modes (this will be the subject of future work). In other side
it seems that the flux tube model \cite{Flux-0} and the QCD sum rules
\cite{sum-decay} predict a suppression of $1^{-+}\textrm{hybrid}\rightarrow\eta\pi$.
This is confirmed using a quite independent model way without any
further hypothesis than the quenched approximation \cite{orsay-selectionRule}.
However, this approximate selection rule is related only to the ``magnetic''
or TE-gluon mode.

\subsection{\label{sub:Results-and-discussion}Results and discussion}

\subsubsection{The mass results}

Our results related to the 1$^{-+}$ hybrid masses and decay widths
for $m_{g}=0.4-0.6$ GeV are summarised in Tables \ref{tab:results1}-\ref{tab:results2},
we add the Table \ref{tab:comparison.} for comparison purposes.

It is difficult to get a hybrid masse lower than $1.5$ GeV ($\gtrsim1.52$
GeV for $m_{g}\gtrsim0.$ ).

We observe a large mixing beween the two QE and GE hybrid modes where
all the TE, TM and longitudinal gluon modes are included in the hybrid
wavefunction (Eq. \ref{eq:wavefunction-long-TE-TM}); for a pure GE-mode
(with \textit{excited} glue $l_{g}=1$ and an S-wave $q\bar{q}$)
we have $M_{1^{-+}}^{\mathrm{GE}}\simeq1.76$$\pm0.05$ GeV for $m_{g}=0.5\pm0.1$
GeV.

Our calculated mass is:
\begin{equation}
M_{1^{-+}}\simeq1.65_{-0.04}^{+0.05}\,\textrm{GeV,}
\end{equation}
which is very close to the latest PDG average \cite{PDG2018}: 
\begin{equation}
1.660_{-0.011}^{+0.015}\,\textrm{GeV,}
\end{equation}

and quite far from $\sim2.$ GeV emerged from the lattice QCD \cite{LQCD-mass}
and the flux tube \cite{Flux-1} studies that systematically discard
the QE-mode where the gluon is not excited i.e. ignore states electric-TM
($j_{g}=1$) and longitudinal (Eq. \ref{eq:QE}). In addition, there
is some difficulties that taint the lattice masse calculations:
\begin{itemize}
\item how to identify interpolation fields used as a hybrid and distinguish
them from ordinary mesons? from the criteria for hybrids proposed
in \cite{LQCD-3} (and adopted implicitely by earlier lattice works
\cite{LQCD-0,LQCD-1,LQCD-2}) the \textit{hybrid-like} character is
directly related to the overlap with the appropriate $J^{\textrm{PC}}$
interpolating fields. This is not always true, we can not understand
the nature of a state by the appearance of its interpolation field.
This is sufficiently illustrated by the strong projection on $\eta$
and $\eta'$ produced with the glue interpolation field $G_{\mu\nu}\widetilde{G}_{\mu\nu}$,
it does not mean that they are glueballs \cite{anti-Dudeck};
\item in the light sector, lattice authors report only results ralted to
the $j_{g}^{\textrm{P}_{g}\textrm{C}_{g}}=1^{+-}$ TE-gluon since
it has the best signal with smallest statistical errors while the
explicit masses of the $j_{g}^{\textrm{P}_{g}\textrm{C}_{g}}=1^{--}$TM-gluon
are not yet published;
\item the lattice calculation sill uses an unrealistic mass of the $\pi$
meson ($\sim396$ MeV) which is much greater than the observed one
($\sim139$ MeV).
\end{itemize}

\subsubsection{The decay results}

Despite imperfections of the model, our predictions are mostly in
reasonable accord with the observed $1^{-+}$ resonance $\pi_{1}(1600)$
seen by several collaborations as shown in Table \ref{tab:comparison.}.
This is espetially true for the controversial $\rho\pi$ channel which
is forbidden by the gluonic excitation models (the $"L+S\,"$ selection
rule \cite{Flux-1-SS-2}). In the constituent glue model the non vanishing
width come from the QE-hybrid mode ($l_{g}=0$ whith $P$-wave $q\bar{q}$,
Eq. \ref{eq:QE}) decaying preferably into two S-wave mesons i.e.:
\begin{equation}
\Gamma_{1^{-+}\rightarrow\rho\pi}\simeq0.28_{-0.09}^{-0.14}\,\textrm{GeV.}
\end{equation}

Our decay results does not take into account the uncertainty due to
$\alpha_{s}$, only errors induced by the $m_{g}$ uncertainties are
considered.

\section{Conclusion}

To conclude, we note that despite the imperfections of the model the
results obtained are encouraging and describe quite well the observed
properties of the resonance $\pi_{1}(1600)$ supporting the fact that
this resonance is a hybrid meson with the internal structure suggested
by the generalized Quark Model with Constituent Gluon i.e. a pair
of quark-antiquar with a massive constituent gluon: $m_{g}\simeq0.5\pm0.1$
GeV. However, this approximate model needs to be improved by considering
more relativistic effects especially for the decay model. On the other
hand, it would be advisable to seriously review the hypothesis that
hybrids are exclusively build by excited gluon fields.
\begin{acknowledgments}
We are grateful to Professor F. Iddir for her help and valuable advice\textbf{.
}Work supported by: PRFU research program (under No. B00L02EP310220190001). \end{acknowledgments}

\begin{table}
\begin{tabular}{|c||c|c|c|c|}
\cline{2-5} 
\multicolumn{1}{c||}{} & $\rho\pi$ & $b_{1}\pi$ & $f_{1}(1235)\pi$ & $\eta'\pi$\tabularnewline
\hline 
\hline 
VES & $\begin{array}{c}
\sim1.6\\
\lesssim0.3*0.34=0.10
\end{array}$\footnote{An experimental decay width limit.} & $\begin{array}{c}
1.56\pm0.06\\
0.34\pm0.06
\end{array}$ & $\begin{array}{c}
1.64\pm0.03\\
0.24\pm0.06
\end{array}$ & $\begin{array}{c}
1.56\pm0.06\\
0.34\pm0.06
\end{array}$\tabularnewline
\hline 
E852 & $\begin{array}{c}
1.593\pm0.008_{-0.047}^{+0.029}\\
0.168\pm0.020_{-0.012}^{+0.150}
\end{array}$\footnote{This state is excluded by a more refined E852 analysis, Ref. \cite{E852-Dzierba}}  & $\begin{array}{c}
1.664\pm0.008\pm0.010\\
0.185\pm0.025\pm0.028
\end{array}$ & $\begin{array}{c}
1.709\pm0.024\pm0.041\\
0.403\pm0.080\pm0.115
\end{array}$ & $\begin{array}{c}
1.597\pm0.010_{-0.010}^{+0.045}\\
0.340\pm0.040\pm0.060
\end{array}$\tabularnewline
\hline 
COMPASS & $\begin{array}{c}
1.600_{-0.060}^{+0.100}\\
0.580_{-0.230}^{+0.100}
\end{array}$ & - & - & $\begin{array}{c}
1.564\pm0.024\pm0.086\\
0.492\pm0.054\pm0.102
\end{array}$\footnote{From JLAC collaboration \cite{JPAC-c} using COMPASS data \cite{COMPASS-3}.}\tabularnewline
\hline 
CLEO & - & - & - & $\begin{array}{c}
1.670\text{\textpm0.0}30\text{\textpm0.0}20\\
0.240\text{\textpm}0.050\text{\textpm0.0}60
\end{array}$\tabularnewline
\hline 
Crystal Barrel & - & $\begin{array}{c}
\sim1.6\\
seen
\end{array}$ & - & -\tabularnewline
\hline 
\end{tabular}\caption{\label{tab:exp}The $\pi_{1}(1600)$ as seen in different experiments.
The resonance masses and the corresponding decay widths are reported
in GeV. }
\end{table}
\begin{table}
\centering{}%
\begin{tabular}{|c|c|c|c|c|c|c|c|}
\hline 
$l_{q\bar{q}}$ & $S_{q\bar{q}}$ & $l_{g}$ & $j_{g}$ & $L$ & hybrid mode & gluon mode & preffered decay mode\tabularnewline
\hline 
\hline 
0 & 1 & 1 & 0 & 0 & GE & long. & $\lyxmathsym{“}L+S\,\lyxmathsym{”}$\tabularnewline
\hline 
0 & 1 & 1 & 1 & 1 & GE & long., TM, TE & $\lyxmathsym{“}L+S\,\lyxmathsym{”}$\tabularnewline
\hline 
0 & 1 & 1 & 2 & 2 & GE & long., TM, TE & $\lyxmathsym{“}L+S\,\lyxmathsym{”}$\tabularnewline
\hline 
1 & 0 & 0 & 1 & 1 & QE & long., TM & $\lyxmathsym{“}S+S\,\lyxmathsym{”}$\tabularnewline
\hline 
\end{tabular}\caption{\label{tab:PC}The lowest $J^{\textrm{PC}}=1^{-+}$ hybrid meson quantum
numbers. Modes and the decay selection rules are shown.}
\end{table}
\begin{table}
\begin{tabular}{|c|c|c|c|}
\hline 
\multicolumn{4}{|c|}{Constituent Glue Model (this work)}\tabularnewline
\hline 
$m_{g}$ & 0.4 & \textbf{0.5} & 0.6\tabularnewline
\hline 
\hline 
$M_{1^{-+}}$ & 1.61 & \textbf{1.65} & 1.70\tabularnewline
\hline 
\hline 
$\Gamma(1^{-+}\rightarrow\rho\pi)$  & 0.19 & \textbf{0.28} & 0.42\tabularnewline
\hline 
$\Gamma(1^{-+}\rightarrow b_{1}\pi)$  & 0.19 & \textbf{0.29} & 0.46\tabularnewline
\hline 
$\Gamma(1^{-+}\rightarrow f_{1}(1235)\pi)$ & 0.06 & \textbf{0.10} & 0.17\tabularnewline
\hline 
\end{tabular}

\caption{\label{tab:results1}The $1^{-+}$ hybrid masses and the decay widths
(in GeV) from the MQGC model.}
\end{table}
\begin{table}
\begin{tabular}{|c|c|c|c||c|c|c|}
\hline 
\multicolumn{4}{|c||}{Constituent Glue Model (this work)} & \multicolumn{2}{c|}{Flux Tube Model} & \multicolumn{1}{c|}{Lattice QCD}\tabularnewline
\hline 
$m_{g}$=0.5 GeV & QE-GE mix.  & QE & GE & Ref.\cite{Flux-1} & Ref.\cite{Flux-2} & Ref. \cite{LQCD-decay}\tabularnewline
\hline 
$\Gamma(1^{-+}hybrid\rightarrow\rho\pi)$  & 0.28 & 0.28 & forbidden \footnote{Selection rule.} & 0.005-0.020 & 0.009 & -\tabularnewline
\hline 
$\Gamma(1^{-+}hybrid\rightarrow b_{1}\pi)$  & 0.29 & forbidden \footnote{Selection rule.} & 0.29 & 0.170 & 0.024 & $0.40\pm0.12$\tabularnewline
\hline 
$\Gamma(1^{-+}hybrid\rightarrow f_{1}(1235)\pi)$ & 0.10 & forbidden \footnote{Selection rule.} & 0.10 & 0.060 & 0.005 & $0.09\pm0.06$\tabularnewline
\hline 
$\Gamma(1^{-+}hybrid\rightarrow\eta\pi,\eta'\pi)$  & - & forbidden \footnote{Allowed beyond the non relativistic approximation. Calculations of
this contribution will be the subject of future work.} & <0.010 & 0.000-0.010 & 0.000 & -\tabularnewline
\hline 
\end{tabular}\caption{\label{tab:results2}Our $1^{-+}$ hybrid masses and the decay widths
(in GeV) compared to the lattice QCD and flux tube results.}
\end{table}
\begin{table}
\begin{tabular}{|c|c||c|c|c|c||c|c|}
\hline 
\multicolumn{2}{|c||}{This work} & \multicolumn{4}{c||}{The experience} & \multicolumn{2}{c|}{Gluonic excitation models}\tabularnewline
\hline 
$m_{g}$ & 0.5$\pm0.1$ & VES & E852 & CLEO & COMPASS & Lattice QCD & Flux tube \footnote{Ref. \cite{Flux-1}.}\tabularnewline
\hline 
\hline 
$M_{1^{-+}}$ & $1.65_{-0.04}^{+0.05}$ & $1.64\pm0.03$ & $1.66\pm0.01$ & $1.67\pm0.04$ & $1.60{}_{-0.06}^{+0.10}$ & $1.79\pm0.14$ \footnote{Ref. \cite{LQCD-mass}.}, $1.96\pm0.04$
\footnote{From Fig. 5 of Ref. \cite{LQCD-3} at $m_{\pi}=396$ MeV ($\ggg139$
MeV, the observed value).} & $\sim2.$00\tabularnewline
\hline 
\hline 
$\rho\pi$  & $0.28_{-0.09}^{+0.14}$ & $\lesssim0.10$ \footnote{An experimental limit.} & $0.17\pm0.02{}_{-0.01}^{+0.15}$ & - & $0.58{}_{-0.23}^{+0.10}$ & - & $0.005-0.020$\tabularnewline
\hline 
$b_{1}\pi$  & $0.29_{-0.11}^{+0.16}$ & $0.34\pm0.06$ & $0.19\pm0.04$ & - & - & $0.40\pm0.12$ \footnote{Ref. \cite{LQCD-decay}} & $0.170$\tabularnewline
\hline 
$f_{1}(1235)\pi$ & $0.10_{-0.04}^{+0.07}$ & $0.24\pm0.06$ & $0.40\pm0.14$ & - & - & $0.09\pm0.06$ \footnote{Ref. \cite{LQCD-decay}} & $0.060$\tabularnewline
\hline 
\end{tabular}\caption{\label{tab:comparison.}The comparison of our work with data and the
recent gluonic excitation models.}
\end{table}

\end{document}